\documentclass[
prc,%
10pt,%
final,%
notitlepage,%
oneside,%
twocolumn,%
nobibnotes,%
nofootinbib,
superscriptaddress,%
floatfix,%
floatfix,%
showkeys,%
showpacs]%
{revtex4}
\usepackage{color}
\usepackage{amsfonts}
\usepackage{amsbsy}
\usepackage{mathrsfs}
\usepackage{graphicx}
\def\lsim{\mathrel{\rlap{
\lower4pt\hbox{\hskip-3pt$\sim$}}
    \raise1pt\hbox{$<$}}}     
\def\gsim{\mathrel{\rlap{
\lower4pt\hbox{\hskip-3pt$\sim$}}
    \raise1pt\hbox{$>$}}}     
\def\scr#1{\mbox{\scriptsize #1}}
\begin{document}
\title{
Vortex rings in fragmentation regions in heavy-ion collisions at $\sqrt{s_{NN}}=$ 39 GeV} 
\author{Yu. B. Ivanov}\thanks{e-mail: Y.Ivanov@gsi.de}
\affiliation{National Research Centre "Kurchatov Institute",  Moscow 123182, Russia} 
\affiliation{National Research Nuclear University "MEPhI", 
Moscow 115409, Russia}
\affiliation{Bogoliubov Laboratory of Theoretical Physics, JINR, Dubna 141980, Russia}
\author{A. A. Soldatov}
\affiliation{National Research Nuclear University "MEPhI",
Moscow 115409, Russia}
\begin{abstract}
Vorticity generated in  heavy-ion collisions at energy of $\sqrt{s_{NN}}=$ 39 GeV
is studied. 
Simulations are performed within a model of the three-fluid dynamics.
A peculiar structure consisting of two vortex rings is found: 
one ring in the target fragmentation region and another one in 
the projectile fragmentation region. 
These rings are also formed in central collisions. 
The matter rotation is opposite in this two rings. 
These vortex rings are already formed at the early stage of the collision together with 
primordial fragmentation regions.  
The average vorticity,  responsible for the global polarization of 
the observed $\Lambda$ and $\bar{\Lambda}$, is also studied. 
In the semi-central collisions
the average vorticity in the midrapidity region turns out to be more than 
an order of magnitude lower than the total one. 
The total vorticity is dominated by the contributions of 
the fragmentation regions and is produced because of asymmetry of the vortex rings
 in noncentral collisions.
This suggests that in semi-central collisions the global polarization 
in the fragmentation regions should be at 
least an order of magnitude higher than that observed by the STAR collaboration 
in the midrapidity. This polarization should be asymmetrical in the reaction plain and  
correlate with the corresponding directed flow. 
\end{abstract}
\maketitle

\section{Introduction}

In peripheral  collisions of high-energy heavy ions the system
has a large angular momentum that results in observable consequences.  
In particular, it was proposed that $\Lambda$ hyperons can be
polarized along the angular momentum of two colliding nuclei \cite{Liang:2004ph,Betz:2007kg,Gao:2007bc}. 
Global polarization of $\Lambda$ and $\bar{\Lambda}$ hyperons  produced in heavy-ion collisions 
has been recently observed \cite{STAR:2017ckg} by the STAR experiment in energy range of 
the Beam Energy Scan (BES) program at the at Relativistic Heavy Ion Collider (RHIC)
at Brookhaven. It was measured in the midrapidity region of colliding 
nuclei. 
It was concluded that the observed rotational fluid has the largest vorticity,
of the order of $\sim$10$^{22}$ s$^{-1}$, that ever existed in the universe.
The measured polarization quantitatively
agrees  with the hydrodynamic \cite{Karpenko:2016jyx,Xie:2017upb} and 
kinetic \cite{Baznat:2017jfj,Baznat:2017ars,Kolomeitsev:2018svb,Li:2017slc}
model calculations.

However, important questions still remain: Why does the global polarization decrease with the collision energy rise
as the total angular momentum increases? Where is the huge angular momentum mainly accumulated? 
These questions were indirectly answered in Refs. \cite{Ivanov:2017dff,Baznat:2015eca,Baznat:2013zx}.   
It was found that the vorticity, which is the driving force of the 
hadron polarization,  is predominantly localized in a relatively thin layer 
at the boundary between participants and spectators.  
This implies that the hyperon polarization should be stronger at peripheral rapidities,  
corresponding to this border than in the midrapidity region. 
This  enhancement at peripheral rapidities was illustrated in Ref. \cite{Baznat:2017ars}
at the example of Au+Au collision at $\sqrt{s_{NN}}=$ 5 GeV. 
Although not all models predict the polarization
minimum at the midrapidity 
\cite{Sun:2017xhx}. 
The polarization enhancement at peripheral rapidities is not a direct 
explanation of the reduction of the midrapidity polarization with the collision energy rise but 
indicates that the hyperon polarization at peripheral rapidities may be very large as compared 
with that at the midrapidity in ultrarelativistic collisions and even higher than that 
at the lowest BEC-RHIC energy.

The study of the hyperon polarization at peripheral rapidities is not just of academic interest. 
Recent proposal \cite{Brodsky:2012vg} to perform experiments at the Large Hadron Collider (LHC) at CERN in the
fixed-target mode revived interest to 
the fragmentation regions in relativistic nucleus-nucleus collisions.
The LHC beam  of lead ions
interacting on a fixed target would provide an opportunity
to carry out precision measurements in the kinematical 
range of the target fragmentation region. 
If the LHC operates in a fixed-target mode at the beam
energy of 2.76 GeV per nucleon for lead nuclei, this is 
equivalent to $\sqrt{s_{NN}}=$ 72 GeV in terms the center-of-mass
energy.

In the present paper we analyze the vorticity in Au+Au collisions at the energy of $\sqrt{s_{NN}}=$ 39 GeV.    
A special emphasis is made on the vorticity in the fragmentation regions. 
The vorticity is simulated within the model of the three-fluid dynamics (3FD) \cite{3FD}.  
The 3FD model is quite successful in reproduction of the major part of bulk
observables: the baryon stopping \cite{Ivanov:2013wha,Ivanov:2012bh}, 
yields of different hadrons, their rapidity and transverse momentum
distributions \cite{Ivanov:2013yqa,Ivanov:2013yla}, and also  
the elliptic \cite{Ivanov:2014zqa} 
and directed \cite{Konchakovski:2014gda} flow excitation functions. 
In recent paper \cite{Ivanov:2018vpw} it was demonstrated that the 3FD model 
successfully reproduces the extensive set of bulk observables at midrapidity
recently presented by the STAR Collaboration \cite{Adamczyk:2017iwn}. 
Therefore, the 3FD predictions for the fragmentation regions may be of
interest.

\section{The 3FD Model}
\label{Model}

The 3FD approximation is a minimal way to simulate the early-stage nonequilibrium 
in the colliding nuclei at high incident energies.
In contrast to the conventional hydrodynamics, 
a finite stopping power resulting in a counterstreaming
regime of leading baryon-rich matter at early stage of the nuclear collision 
is taken into account in the 3FD description \cite{3FD}. This 
nonequilibrium state of the baryon-rich matter
is modeled by two interpenetrating baryon-rich fluids 
initially associated with constituent nucleons of the projectile
(p) and target (t) nuclei. 
At later stages these baryon-rich fluids may consist
of any type of hadrons and/or partons (quarks and gluons),
rather than only nucleons.
Newly produced particles, which
populate the midrapidity region, are associated with a fireball
(f) fluid.
Each of these fluids is governed by conventional hydrodynamic equations 
coupled by friction terms in the right-hand sides of the Euler equations. 
These friction terms describe energy--momentum loss of the 
baryon-rich fluids. A part of this
loss is transformed into thermal excitation of these fluids, while another part 
gives rise to particle production into the fireball fluid.

The physical input of the present 3FD calculations is described in
Ref.~\cite{Ivanov:2013wha}. The friction between fluids was fitted for each EoS to reproduce
the observed stopping power, 
see  Ref. \cite{Ivanov:2013wha} for details.
The simulations in 
\cite{Ivanov:2013wha,Ivanov:2012bh,Ivanov:2013yqa,Ivanov:2013yla,Ivanov:2014zqa,Konchakovski:2014gda} 
were performed with different 
equations of state (EoS's)---a purely hadronic EoS \cite{gasEOS}  
and two versions of the EoS involving the   deconfinement
 transition \cite{Toneev06}, i.e. a first-order phase transition  
and a smooth crossover one. In the present paper we demonstrate results with 
only the first-order-phase-transition (1st-order-tr.) and crossover EoS's as 
the most successful in reproduction of various observables, in particular, the recent data on 
bulk observables at $\sqrt{s_{NN}}=$ 39 GeV \cite{Ivanov:2018vpw}.

In semi-central Au+Au collisions 
at the energy of $\sqrt{s_{NN}}=$ 39 GeV  the counterstreaming
of leading baryon-rich matter takes place only during short initial 
stage. After a short time $\approx$1 fm/c 
the p- and t-fluids are
either spatially separated or mutually stopped and unified, 
thus forming a unified single baryon-rich fluid 
similarly to that happening in central collisions at this energy \cite{Ivanov:2017xee}. 
Indeed, the corresponding unification measure 
   \begin{eqnarray}
   \label{unification}
   1-\frac{n_{\rm p}+n_{\rm p}}{n_{\rm B}}
   \end{eqnarray}
at $t\gsim 1$ fm/c is less than 0.01. Here 
   \begin{eqnarray}
   \label{nb-prop}
   n_{\rm B} = |J_{\rm B}|= \left(J_{\rm B}^{\mu} J_{{\rm B}\mu}\right)^{1/2} 
   \end{eqnarray}
is the proper (i.e. in the local rest frame) baryon density of the unified baryon-rich fluid 
defined in terms of the baryon current
$J_{\rm B}^{\mu} = n_{\rm p}u_{\rm p}^{\mu}+n_{\rm t}u_{\rm t}^{\mu}$, where   
$n_{\alpha}$ and
$u_{\alpha}^{\mu}$ are the proper baryon densities and 
 hydrodynamic 4-velocities of the p- and t-fluids, respectively.  
This unification measure is zero, when the p- and t-fluids are mutually stopped and unified, and  
has a positive value increasing  with rise of
the relative velocity of the p- and t-fluids.

The f-fluid also is entrained by the the unified baryon-rich fluid 
but is not that well unified with the latter. The local baryon-fireball relative velocity is
$ v_{fB} \lsim$ 0.2, at $t \gsim$ 1 fm/c, i.e. it is small but not negligible. 
In particular, the friction between
the baryon-rich and net-baryon-free fluids is the
only source of dissipation at the expansion stage. 
Thus, at $t \gsim$ 1 fm/c the system is characterized by two hydrodynamical velocities,  
$u_{\rm B}^{\mu}$ and $u_{\rm f}^{\mu}$,  attributed to the baryon-rich and baryon-free fluids, respectively, 
and two corresponding temperatures $T_{\rm B}$ and $T_{\rm f}$.

\begin{figure}[htb]
\includegraphics[width=8.2cm]{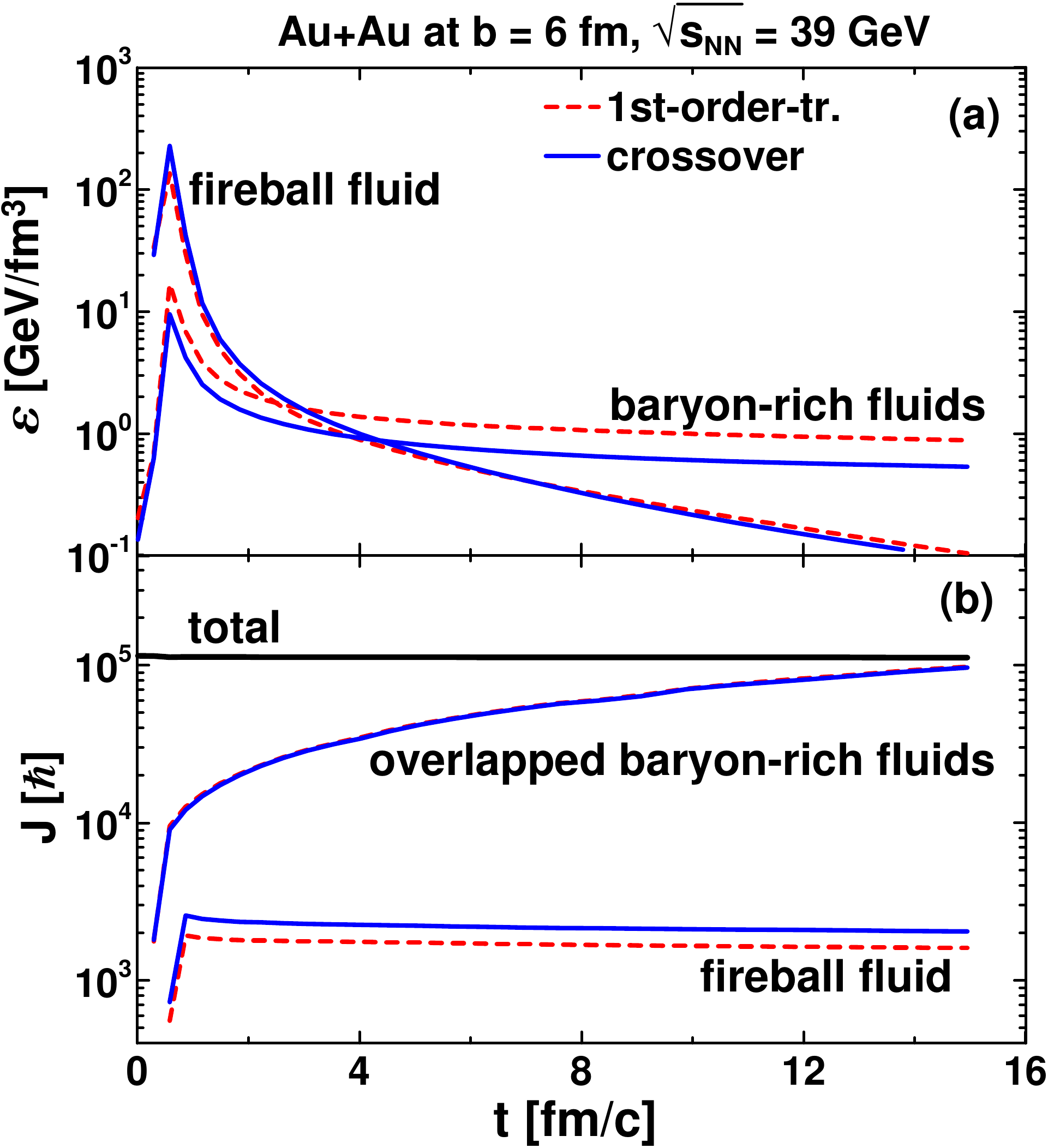}
 \caption{(Color online)
Time evolution of 
(a) average proper energy densities of the baryon-rich and baryon-free fluids
and 
(b) the total angular momentum (conserved quantity), 
the angular momentum accumulated in the baryon-rich fluids in their overlap 
region and the angular momentum of the fireball fluid 
in the semi-central ($b=$ 6 fm) Au+Au collision at $\sqrt{s_{NN}}=$ 39 GeV. 
Calculations are done with the first-order-phase-transition and crossover EoS's. 
}
\label{fig0}
\end{figure}

In Fig. \ref{fig0} the time evolution of 
average proper energy densities of the baryon-rich and baryon-free fluids [panel (a)]
and the total angular momentum, 
the angular momentum of the baryon-rich fluids in their overlap 
region and the angular momentum of the fireball fluid [panel (b)]
in the semi-central ($b=$ 6 fm) Au+Au collision at $\sqrt{s_{NN}}=$ 39 GeV are presented. 
As seen, at $t<$ 4 fm/c the f-fluid dominates, i.e. its average energy density exceeds 
that of the baryon-rich fluid. Whereas at $t>$ 4 fm/c, 
the situation is reversed -- the baryon-rich fluid dominates. As seen from  Fig. \ref{fig1}
(baryon-density and energy-density columns), 
it dominates in fragmentation regions where the baryon charge is concentrated.  

The total angular momentum (that includes a contribution of spectators) is a conserved quantity. 
It is determined as 
   \begin{eqnarray}
   \label{J-tot}
J = \int d^3 x \sum_{\alpha=\rm{p,t,f}} (z\; T^{\alpha}_{10} - x\; T^{\alpha}_{30}).    
   \end{eqnarray}
where $T^{\alpha}_{\mu\nu}$ is the energy-momentum tensor of the $\alpha$(=p,t,f) fluid that 
have the conventional hydrodynamical form, $z$ is the beam axis, 
$(x,z)$ is the reaction plane of the colliding nuclei. 
The constancy of the total angular momentum [Fig. \ref{fig0}, panel (b)]
demonstrates the accuracy of the numeric scheme: $J_{\rm{total}}$
is conserved with the accuracy of 1\%. The overlap region of the baryon-rich fluids
rises in the course of the expansion stage and includes 
more and more former spectators.  
Thus, the angular momentum of the baryon-rich fluid in their overlap 
almost accumulates the major part of the total angular momentum of the system at the final stage of the collision, 
cf. Fig. \ref{fig0}, panel (b). 
The angular momentum of the newly produced f-fluid is almost 
two orders of magnitude lower than that of the overlapped baryon-rich fluids at the final stage of the 
collision.

\section{Vorticity in the 3FD model}
\label{Results}

There are several definitions of the vorticity that are suitable 
for calculations of the hadron polarization in different approaches. In the present study we  
consider the relativistic kinematic vorticity
   \begin{eqnarray}
   \label{rel.kin.vort.}
   \omega_{\mu\nu} = \frac{1}{2}
   (\partial_{\nu} u_{\mu} - \partial_{\mu} u_{\nu}), 
   \end{eqnarray}
where $u_{\mu}$ is a collective local four-velocity of the matter. 
This type of the vorticity is directly relevant to the hadron polarization due to 
the chiral vortical effect \cite{Sorin:2016smp}
that was used in Refs. \cite{Baznat:2017jfj,Baznat:2017ars}. 
Another type of the vorticity is so-called thermal vorticity 
   \begin{eqnarray}
   \label{therm.vort.}
   \varpi_{\mu\nu} = \frac{1}{2}
   (\partial_{\nu} \hat{\beta}_{\mu} - \partial_{\mu} \hat{\beta}_{\nu}), 
   \end{eqnarray}
where $\hat{\beta}_{\mu}=\hbar\beta_{\mu}$ and $\beta_{\mu}=u_{\nu}/T$ 
with $T$ being the local temperature. Thus, $\varpi$ is dimensionless. 
It is directly related to the polarization vector of a spin 1/2 particle
in the thermodynamical approach \cite{Becattini:2013fla} that was used in Refs.   
\cite{Karpenko:2016jyx,Xie:2017upb,Kolomeitsev:2018svb,Li:2017slc}.

As it was argued above, at $t \gsim$ 1 fm/c the system is characterized by two hydrodynamical velocities,  
$u_{\rm B}^{\mu}$ and $u_{\rm f}^{\mu}$, and two corresponding temperatures, $T_{\rm B}$ and $T_{\rm f}$, 
corresponding to the unified baryon-rich fluid (B) and the baryon-free fluid (f).  
Thus we deal with two sets of the vorticity attributed  to  
to these baryon-rich and baryon-free fluids, respectively. 
%
%

In order to suppress contributions of almost empty regions, 
where the matter is relatively thin, we consider a  
proper-energy-density weighted  vorticity in the reaction ($xz$) plane, 
i.e. at $y=0$, similarly to that done in Ref. \cite{Ivanov:2017dff}.  
Moreover, we sum the vorticity of the baryon-rich and baryon-free fluids 
with the weights of their energy densities in order to define a single 
quantity responsible for the vortical effects, in particular, the particle polarization. 
Thus, the proper-energy-density weighted kinematic vorticity is defined as 
   \begin{eqnarray}
   \label{en.av.rel.B-vort}
   &&\Omega_{\mu\nu} (x,0,z,t) = [\omega_{\mu\nu}^{\rm B}(x,0,z,t)  \varepsilon_{\rm B} (x,0,z,t)
\cr
   &+& \omega_{\mu\nu}^{\rm f}(x,0,z,t)  \varepsilon_{\rm f} (x,0,z,t)]
   / \langle \varepsilon (y=0,t) \rangle , 
   \end{eqnarray}
where $\varepsilon_{\rm B}$ and $\varepsilon_{\rm f}$ are the proper energy densities of the 
the baryon-rich and baryon-free fluids, respectively. 
The total proper energy density of all three fluids in the local rest frame, $\varepsilon$,  
is strictly defined as follows
\begin{eqnarray}
\label{eps_tot}
\varepsilon = u_\mu T^{\mu\nu} u_\nu. 
\end{eqnarray}
in terms of the total energy--momentum tensor
$T^{\mu\nu} \equiv
T^{\mu\nu}_{\scr p} + T^{\mu\nu}_{\scr t} + T^{\mu\nu}_{\scr f}$
%
%
being the sum of conventional hydrodynamical energy--momentum tensors of separate fluids, and
the total collective 4-velocity of the matter
\begin{eqnarray}
\label{u-tot}
u^\mu = u_\nu T^{\mu\nu}/(u_\lambda T^{\lambda\nu} u_\nu). 
\end{eqnarray}
Note that definition (\ref{u-tot}) is, in fact, an equation
determining $u^\mu$. In general, this $u^\mu$ does not coincide with 
4-velocities of separate fluids. 
However, in view of almost perfect unification of the baryon-rich fluids 
and small local baryon-fireball relative velocities, 
$ v_{fB} \lsim$ 0.2, at $t \gsim$ 1 fm/c, 
a very good approximation for $\varepsilon$ is just 
\begin{eqnarray}
\label{eps-tot-appr}
\varepsilon \simeq  \varepsilon_{\scr B} + \varepsilon_{\scr f}. 
\end{eqnarray}
The average energy density in the $xz$ plane in Eq. (\ref{en.av.rel.B-vort}) is
   \begin{eqnarray}
   \label{B-en.av.}
\langle \varepsilon (y,t) \rangle = 
\int dx \; dz \; \varepsilon (x,y,z,t) 
\Big/
\int_{\varepsilon (x,y,z,t)>0} dx \; dz.
   \end{eqnarray}
Similarly to $\Omega_{\mu\nu}$, 
we define the proper-energy-density weighted  thermal vorticity 
in the reaction plane, though keep the same notation ($\varpi_{\mu\nu}$) for it.

The summation of   $\omega_{\mu\nu}^{\rm B}$ and $\omega_{\mu\nu}^{\rm f}$ 
with the weights of their energy densities in Eq.   (\ref{en.av.rel.B-vort})
requires certain comments. Thus defined kinematic vorticity is very 
close to the true kinematic vorticity of the composed matter when the B- and f-velocities are close 
to each other. The latter is indeed the case at $t \gsim$ 1 fm/c. 
Although, for the thermal vorticity  this is not true. 
It is impossible to unambiguously define an effective temperature of  the composed matter. 
Therefore, the energy-density-weighted sum for the thermal vorticity of 
a thermally non-equilibrated system is a certain anzatz giving us a possibility 
to analyze a nonequilibrium system in equilibrium terms. 
Nevertheless, this anzatz is very close to the true thermal vorticity in certain cases. 
As it was mentioned above, the f-fluid energy density is lower than that of the unified 
baryon-rich fluid at $t>$ 4 fm/c. 
Therefore, at $t>$ 4 fm/c the f-fluid can be considered as small perturbation and hence 
this anzatz for the thermal vorticity of the composed matter
is a good approximation.

\begin{figure*}[!htb]
\includegraphics[width=18.cm]{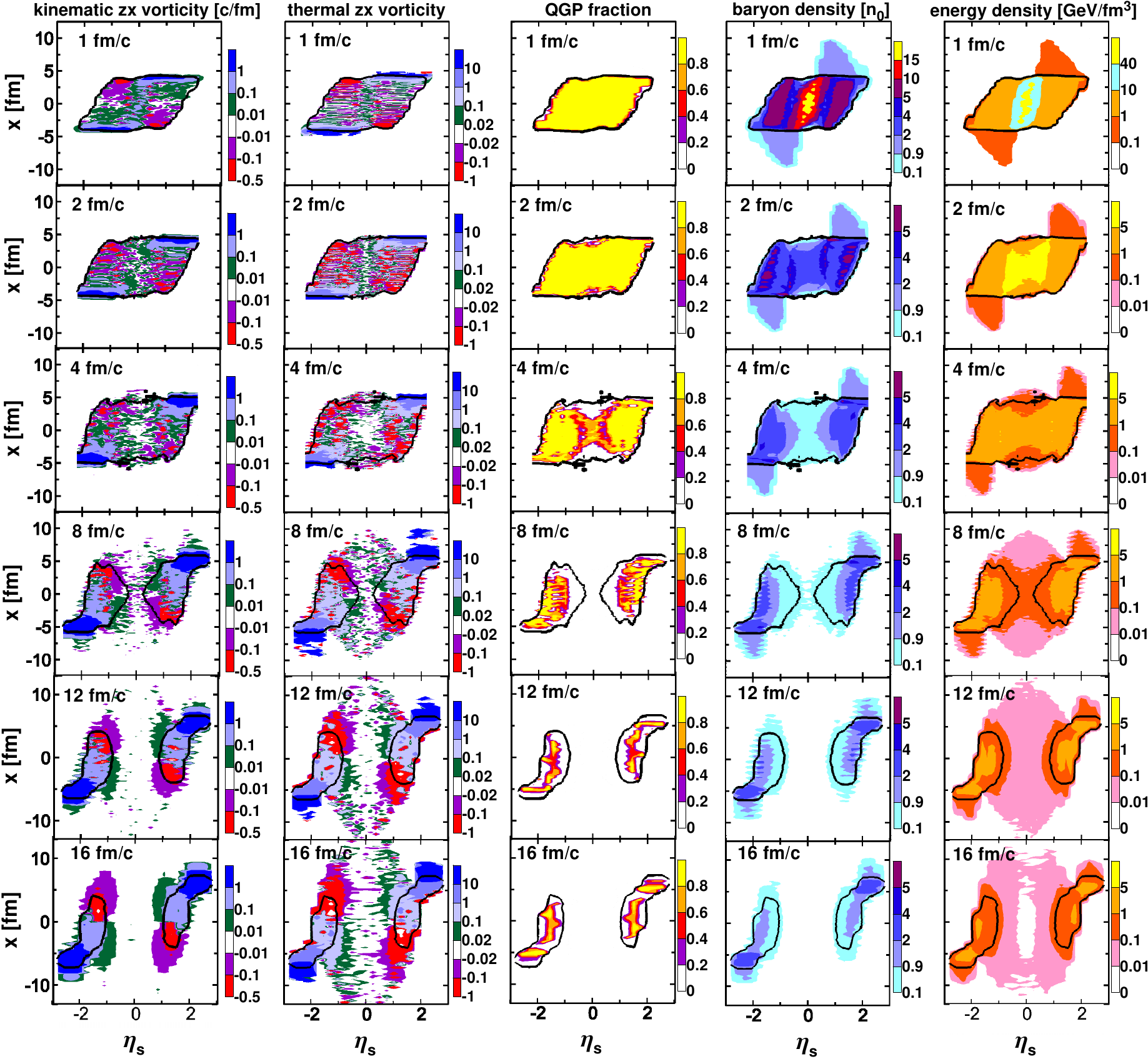}
 \caption{(Color online)
Columns from left to right: The proper-energy-density weighted 
relativistic kinematic $zx$ vorticity, Eq.   (\ref{en.av.rel.B-vort}), 
the similarly weighted thermal $zx$ vorticity, 
the QGP fraction, the proper baryon density ($n_B$)  
[see Eq. (\ref{nb-prop})] in units of the the normal nuclear density ($n_0=0.15$ 1/fm$^3$),
and the proper energy density ($\varepsilon$), see Eq. (\ref{eps_tot}),
in the reaction plane at various time instants 
in the semi-central ($b=$ 6 fm) Au+Au collision at $\sqrt{s_{NN}}=$ 39 GeV. 
$\eta_s$  is the space-time rapidity along the beam direction, see Eq. (\ref{eta_s}).
Calculations are done with the first-order-phase-transition EoS. $z$ axis is the 
beam direction.  
The bold solid contour displays the border of the frozen out matter.  Inside this contour 
the matter still hydrodynamically evolves, while outside -- it is frozen out.
}
\label{fig1}
\end{figure*}

The 3FD simulations of Au+Au collisions were performed 
without freeze-out. 
The freeze-out in the 3FD model removes the frozen out matter from the hydrodynamical 
evolution \cite{Russkikh:2006aa}. Therefore, in order to keep all the matter in the 
consideration the freeze-out was turned off.

\begin{figure}[!tbh]
\includegraphics[width=8.6cm]{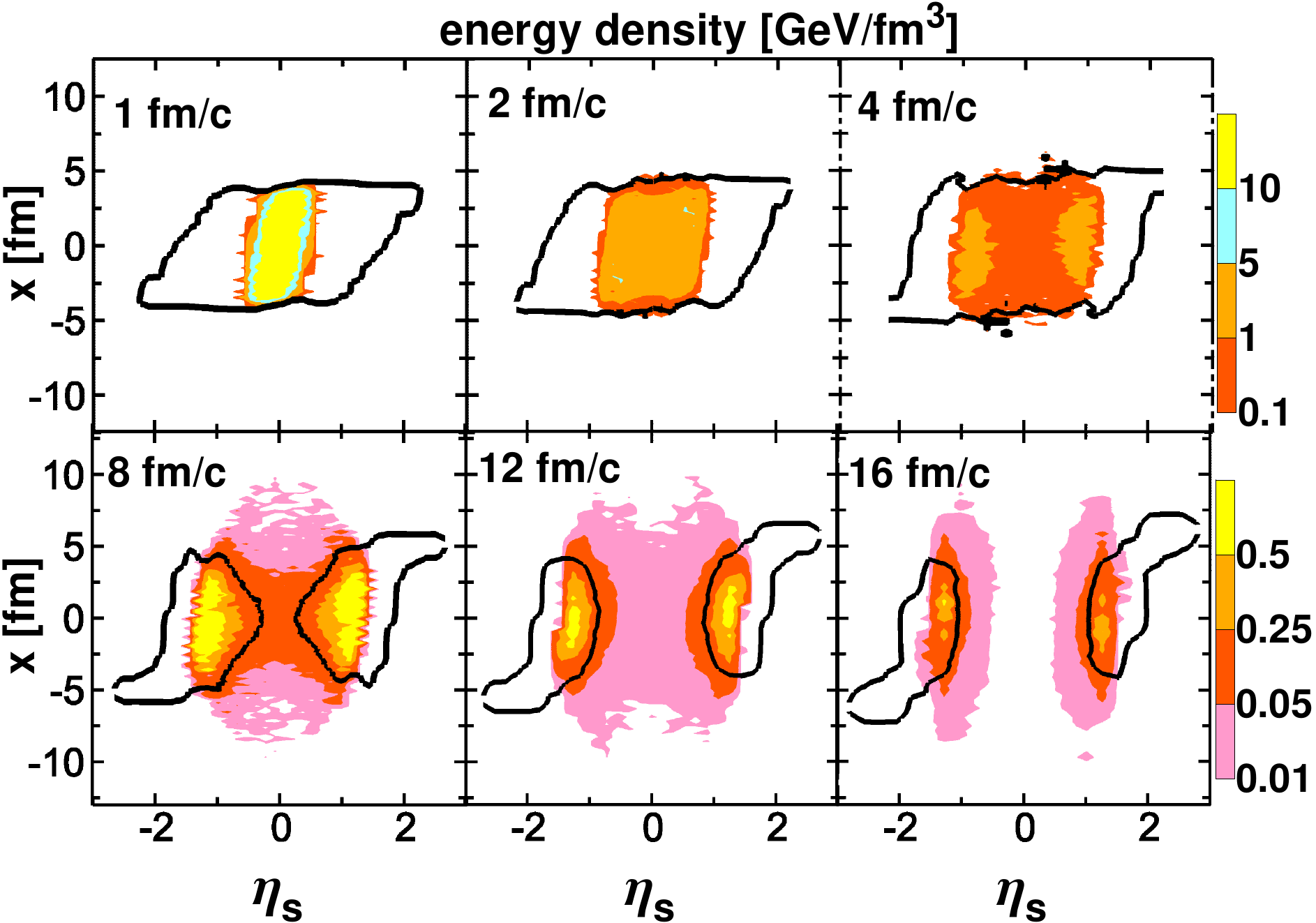}
 \caption{(Color online)
The proper energy density of the f-fluid 
in the reaction plane at various time instants 
in the semi-central ($b=$ 6 fm) Au+Au collision at $\sqrt{s_{NN}}=$ 39 GeV. 
$\eta_s$  is the space-time rapidity along the beam direction.
Calculations are done with the first-order-phase-transition EoS. $z$ axis is the 
beam direction.  
The bold contours display the borders between the frozen-out
and still hydrodynamically evolving matter.
}
\label{fig2}
\end{figure}

Figure \ref{fig1} presents the time evolution of the proper-energy-density weighted 
relativistic kinematic $zx$ vorticity of the composed matter, see Eq. (\ref{en.av.rel.B-vort}), 
the similarly weighted thermal $zx$ vorticity, the QGP fraction, and     
the proper baryon and energy densities, Eqs. (\ref{nb-prop}) and (\ref{eps_tot}), respectively, 
in the reaction plain ($x\eta_s$) of
central Au+Au collision at $\sqrt{s_{NN}}=$ 39 GeV, where 
   \begin{eqnarray}
   \label{eta_s}
   \eta_s = \frac{1}{2} \ln\left(\frac{t+z}{t-z}\right)
   \end{eqnarray}
is the longitudinal space-time rapidity and $z$ is the coordinate along the beam direction. 
The advantage of this longitudinal space-time rapidity is that it is 
equal to the kinematic longitudinal rapidity 
defined in terms of the longitudinal velocity in the self-similar one-dimensional expansion 
of the system.  
As already mentioned, the baryon-rich fluids are  
mutually stopped and unified at $t\gsim 1$ fm/c.  
As the f-fluid is not that well unified with the combined baryon-rich fluid, 
the evolution of the f-fluid is separately presented in Fig. \ref{fig2}. 
The baryon-rich fluid entrains the f-fluid. This is the reason for the smallness of $v_{{\scr f}B}$.

Thus, the dissipation in the system is very moderate at $t\gsim 1$ fm/c because the system 
is almost kinetically equilibrated. This smallness of the dissipation was confirmed by the analysis 
of the entropy production in the 3FD simulations \cite{Ivanov:2016vkw}. Therefore, it may seem that 
an additional conservation law, i.e. the circulation conservation (Kelvin's circulation theorem), may be approximately valid. However, this is not the case. The Kelvin's circulation theorem 
was originally formulated for a so-called barotropic fluid, i.e. the fluid with the EoS of the type of 
$P=P(n_B)$, where $P$ is the pressure. In fact, it is not important that $P$ depends on $n_B$, 
it is important that $P$ depends on a single variable (it can also be the energy density $\varepsilon$). 
In our case we have essentially $P=P(n_B,\varepsilon)$ because the system is essentially baryon-rich, 
as seen from  Fig. \ref{fig1}. 
Therefore, the circulation is not conserved.

All major features of the collision dynamics are similar to those described in Ref. \cite{Ivanov:2017xee}
for the central collision at the same energy. 
As seen from Fig. \ref{fig1}, at $t=$ 1 fm/c
the thermalized central (see the central bumps in $n_B$ and $\varepsilon$) 
and primordial fragmentation regions,
i.e. the baryon-rich matter  passed through the interaction region 
(see two bumps of baryon density near $\eta_s=\pm$ 1), 
have already been formed. 
The matter in all these regions is in the quark-gluon phase, 
see the QGP fraction in Fig. \ref{fig1}.  
There is a large fraction of the baryon charge stopped in 
the central region. This is in contrast 
to the ultra-relativistic  scenario (at the top RHIC and LHC energies) where the major 
part of the baryon charge is assumed to be located in the fragmentation regions
already at the initial stage. 
The proper baryon and energy densities  in this central region approximately are
$n_B/n_0 \approx$ 10 and $\varepsilon\approx$ 10 GeV/fm$^3$, respectively.
The proper baryon density is similar to that attained in the central collision 
\cite{Ivanov:2017xee} while the energy one is considerably lower. 
The f-fluid dominates in this central energy density, as seen from Fig. \ref{fig0}.

In the course of time, the central region undergoes a  
rapid, practically self-similar one-dimensional (1D) expansion. 
In fact, the thermalized central region is produced in the state of this expansion. 
The matter, and in particular the baryon charge, is pushed out to the periphery of this central fireball, 
i.e. closer to the primordial fragmentation regions. 
The primordial fragmentation fireballs also expand in counter directions to the central one. 
The primordial fragmentation fireballs join 
with central contributions to the  
instant $t=4$ fm/c because of their counter expansion, see Fig. \ref{fig1}.   
Therefore, the final fragmentation regions consist of primordial fragmentation fireballs  
and baryon-rich regions of the central fireball pushed out to peripheral rapidities. 
However, complete mixing of these central and primordial fragmentation fireballs   
does not occur, 
as seen from Fig. \ref{fig2}.

At later time $t \geq$ 8 fm/c, see Figs. \ref{fig1} and \ref{fig2}, 
the central part of the system gets frozen out while the fragmentation regions continue 
to evolve being already separated in the configuration space. 
This longer evolution of the fragmentation regions is due to the relativistic time dilation 
caused by their high-speed motion with respect to the central region. 
Therefore, their evolution time in the c.m. frame of colliding nuclei
lasts $\approx 30$ fm/c in the first-order-transition scenario and $\approx 25$ fm/c in the crossover one.

From the very beginning
the vortical fields, both the kinematic and thermal ones, 
are predominately formed at the periphery of the system, 
i.e. at the border between the participant and spectator matter, see Fig. \ref{fig1}.  
This means that the vorticity is initially located at peripheral rapidities rather than 
at midrapidity. 
Later on, the vortical fields partially spread to the participant and spectator bulk though 
remain concentrated near the border.  In the conventional hydrodynamics this 
extension into the bulk is an effect of the shear viscosity. In the 3FD dynamics it 
is driven by the 3FD dissipation which imitates the effect of the shear viscosity \cite{Ivanov:2016vkw}.
The spread into the bulk, i.e. into the midrapidity region, 
is stronger at lower collision energies  \cite{Ivanov:2017dff}
because of the higher effective shear viscosity than that at higher energies \cite{Ivanov:2016vkw}. 
This explains the drop of the vorticity value and consequently 
the observed hyperon polarization at the midrapidity with the collision energy rise.

At later times the maximum values in the vortical fields get more and more shifted 
to the fragmentation regions because of the 1D expansion of the system. 
At the same time, the vorticity in the participant bulk gradually dissolves  
and practically disappears in the center of the colliding system. 
This longer evolution of the fragmentation regions is a result of the 
above mentioned relativistic time dilation 
caused by their high-speed motion with respect to the central region.

It is peculiarly that four strong oppositely directed vortices are formed at the periphery 
of the fragmentation regions, see Fig. \ref{fig1}.   
These strong oppositely directed vortices pedominantely consist 
of the baryon-rich matter because the f-fluid lags behind the primordial fragmentation matter, 
see Fig. \ref{fig2}. 
Over time, these vortices capture ever larger areas.
The vortex at the border with the spectator matter is an order of magnitude stronger 
than its counterpart. This is the structure as it is seen in the reaction plane.

\begin{figure}[!htb]
\includegraphics[width=4.5cm]{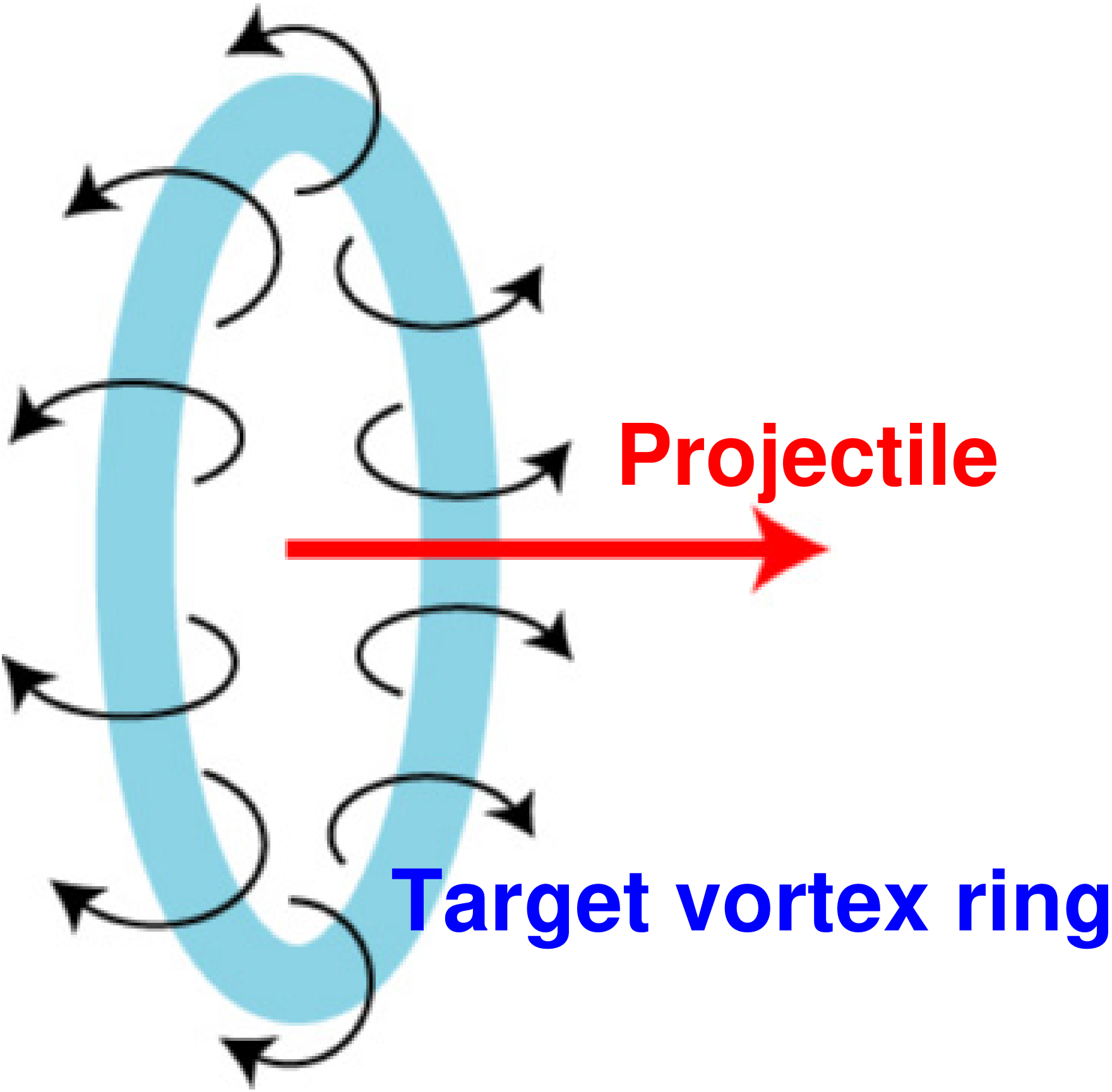}
 \caption{(Color online)
Schematic picture of the vortex ring in the target fragmentation region. 
Curled arrows indicate direction of the circulation of the target matter. 
}

\label{fig3a}
\end{figure}
\begin{figure}[!htb]
\includegraphics[width=8.6cm]{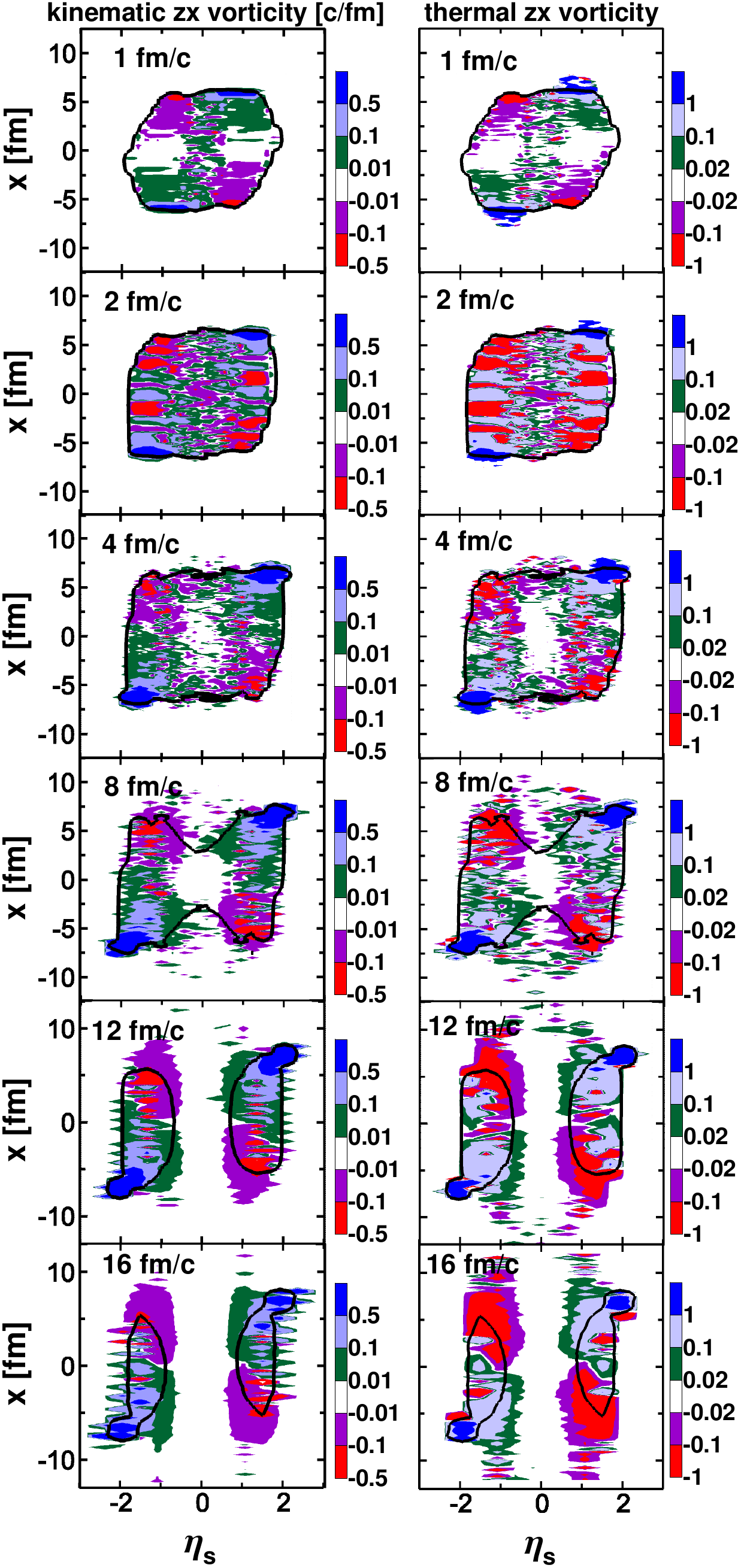}
 \caption{(Color online)
The same as in Fig. \ref{fig1} but for  the central ($b=$ 2 fm) Au+Au collision at $\sqrt{s_{NN}}=$ 39 GeV
and without three right columns.
}
\label{fig3}
\end{figure}

In fact, in three dimensions these are two vortex rings: one in the target fragmentation region and another in 
the projectile one. The matter rotation is opposite in this two rings. They are 
formed because the matter in the vicinity of the beam axis ($z$) is stronger decelerated 
 because of thicker matter in the center than that at the periphery. 
Indeed, these rings are formed at the transverse periphery 
of the stopped matter in the central region, see the central bumps in $n_B$ and $\varepsilon$ 
at $t=$ 1 fm/c in Fig. \ref{fig1}. 
Thus, the peripheral matter acquires a rotational motion. 
A schematic picture of the vortex ring in the target fragmentation region
is presented in Fig. \ref{fig3a}.  

A similar effect was noticed in the analysis of the vorticity field 
\cite{Ivanov:2017dff,Baznat:2015eca,Baznat:2013zx} at energies of 
the  Nuclotron-based Ion Collider fAcility (NICA) at the 
Joint Institute for Nuclear Research (JINR) in Dubna. 
The authors of  Refs. \cite{Baznat:2015eca,Baznat:2013zx} called this 
specific toroidal structure as a femto-vortex sheet. 
This femto-vortex sheet is not a ring because the vorticity disappears 
in the $xy$ plane, i.e. in the plane orthogonal to the reaction $xz$ plane. 
At $\sqrt{s_{NN}}=$ 39 GeV this femto-vortex sheet splits into two real 
rings, in which the vorticity does not disappear in the $xy$ plane 
though is anisotropic in the $x$ direction in noncentral collisions. 

These rings are also formed in central collisions, as seen from Fig. \ref{fig3}. 
In the case of $b=$ 2 fm the distribution of the vorticity along the rings 
is more homogeneous than that at $b=$ 6 fm, but still the vorticity reaches high values. 
As seen from Fig. \ref{fig3}, the vortex rings are already formed at $t=$ 1 fm/c with the vorticity 
predominantly concentrated at the periphery of the fragmentation zone. In the course of time 
the vorticity spreads into the bulk of the fragmentation regions. 
In fact, the  schematic picture of the completely symmetric vortex ring, 
see Fig. \ref{fig3a}, corresponds to the exactly central collision at $b=$ 0.

%
\begin{figure}[bht]
\includegraphics[width=8.4cm]{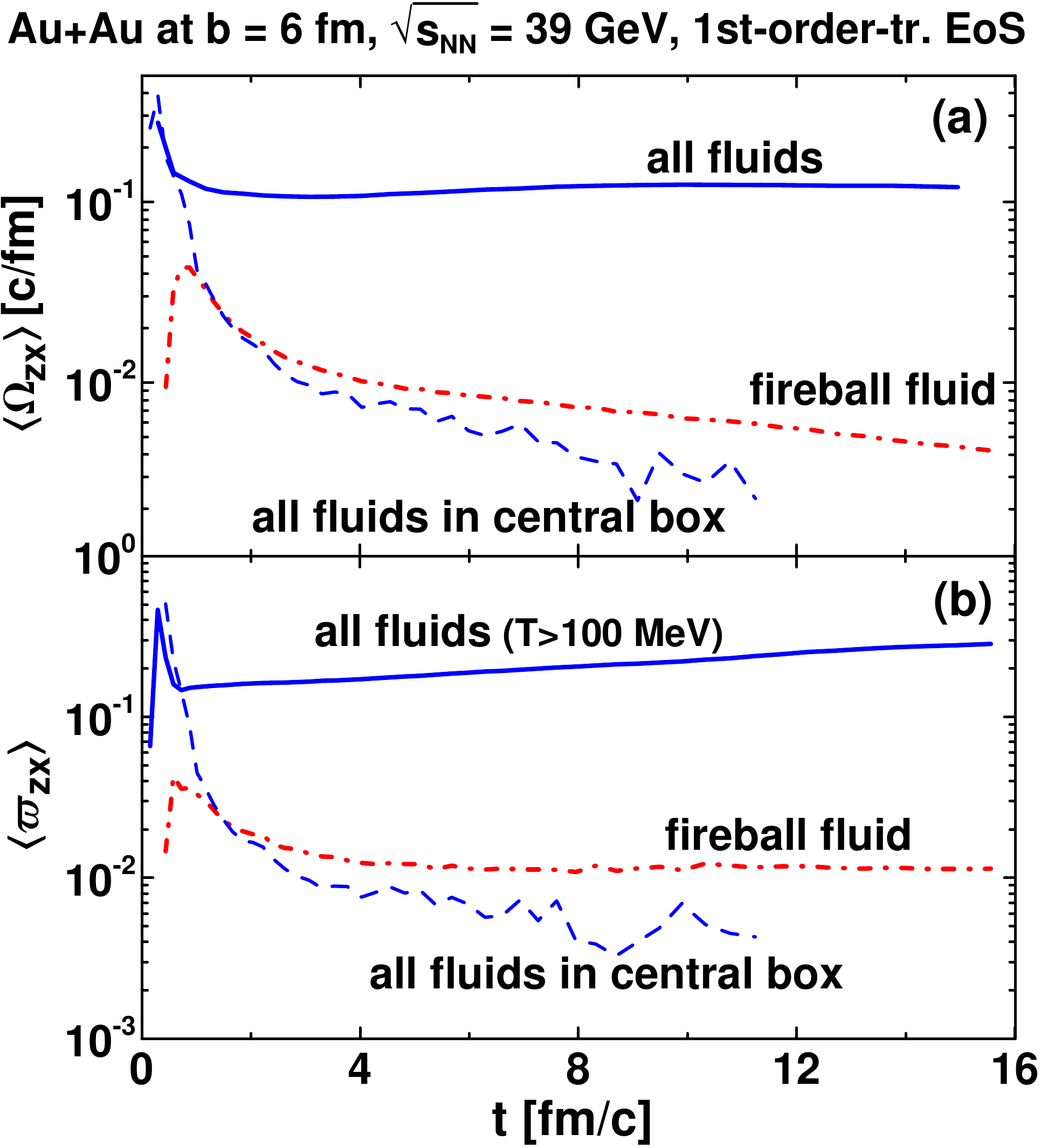}
 \caption{(Color online)
Time evolution of the proper-energy-weighted (a) 
relativistic kinematic $zx$ vorticity and (b) thermal $zx$ vorticity
in the semi-central ($b=$ 6 fm) Au+Au collision at $\sqrt{s_{NN}}=$ 39 GeV. 
The vorticity of the composed matter, Eq. (\ref{en.av.rel.B-vort-T}), 
and that of the f-fluid are averaged over the whole system. 
The thermal vorticity of the composed matter, Eq. (\ref{en.av.therm.B-vort-T}),
is averaged only over regions with high temperature, $T>$ 100 MeV.   
The vorticities in the central box are averaged accordingly to Eqs. (\ref{en.av.rel.B-vort-T})
and (\ref{en.av.therm.B-vort-T}) but only over the central region $|x| < R-b/2$, $|y| < R-b/2$
and $|x| < R/\gamma_{cm}$, 
where $R$ is the radius of the Au nucleus and 
$\gamma_{cm}$ is the Lorentz
factor associated with the initial nuclear motion in the c.m. frame
Calculations are done with the first-order-transition EoS.
}
\label{fig4}
\end{figure}
%
%
\begin{figure}[bht]
\includegraphics[width=8.cm]{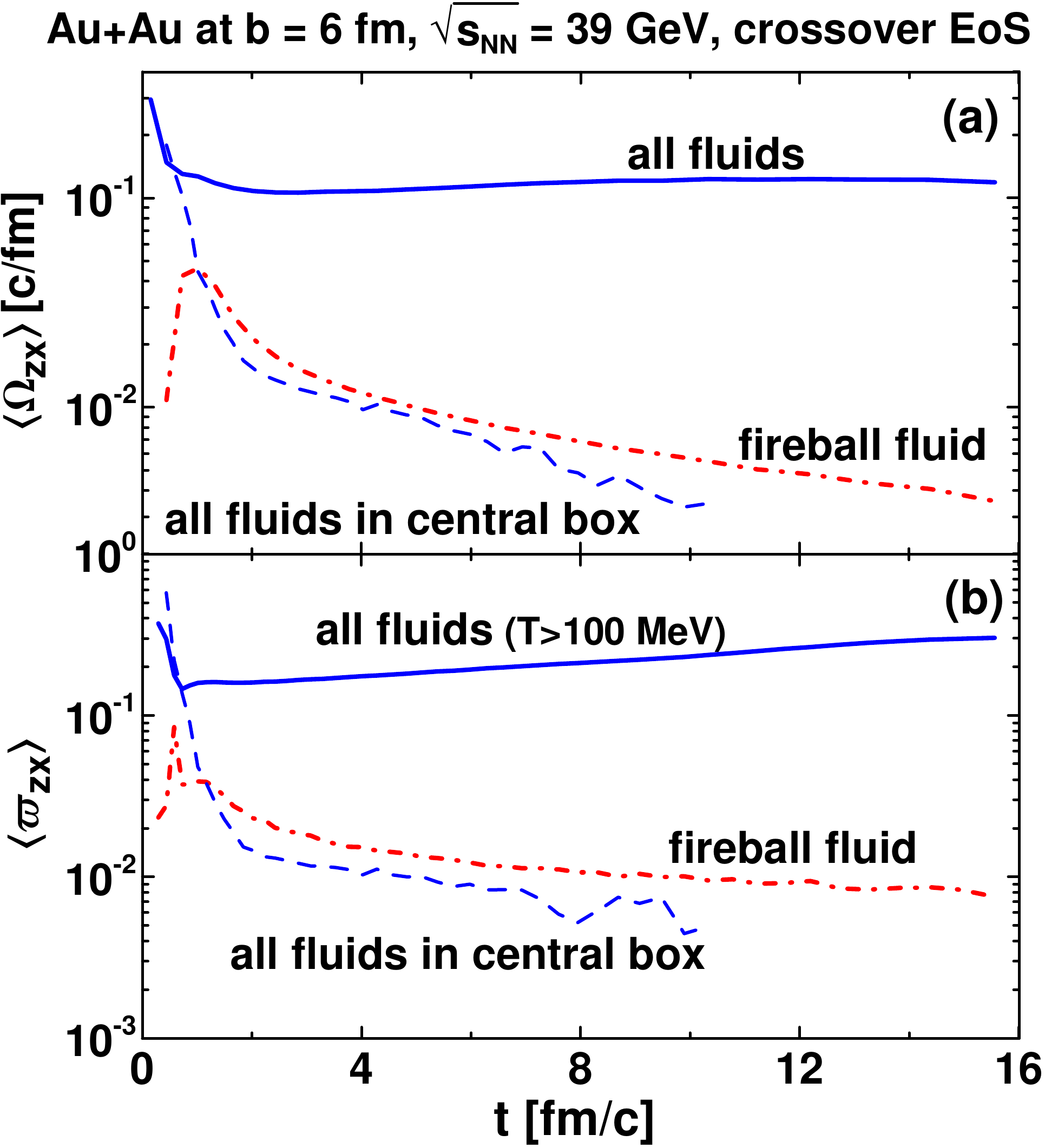}
 \caption{(Color online)
The same as in Fig. \ref{fig4} but for 
the crossover EoS.
}
\label{fig5}
\end{figure}

In order to perform a more quantitative comparison of the vorticity in the
midrapidity and fragmentation regions
we calculated
(kinematic and thermal) $zx$ vorticity  of the composed matter 
[in the sense of the anzatz of Eq.   (\ref{en.av.rel.B-vort})]
in the semi-central ($b=$ 6 fm) Au+Au collision at $\sqrt{s_{NN}}=$ 39 GeV 
averaged with the weight of the proper energy density over the whole system 
   \begin{eqnarray}
   \label{en.av.rel.B-vort-T}
   \langle \Omega_{\mu\nu}(t)\rangle &=& \int dV \;
   [\omega_{\mu\nu}^{\rm B}({\bf x},t)\;\varepsilon_{\rm B}({\bf x},t)
   \cr
   &+&\vphantom{\int dV}\omega_{\mu\nu}^{\rm f}({\bf x},t)\;\varepsilon_{\rm f}({\bf x},t)]
 \Big/ \langle \varepsilon (t)\rangle
   \end{eqnarray}
   \begin{eqnarray}
   \label{en.av.therm.B-vort-T}
  \langle \varpi_{\mu\nu} (t) \rangle_{T>T_0} &=& \int_{T>T_0} dV \;
   [\varpi_{\mu\nu}({\bf x},t)_{\rm B}\;\varepsilon_{\rm B} ({\bf x},t) 
   \cr
   &+&\vphantom{\int dV}\varpi_{\mu\nu}^{\rm f}({\bf x},t)\;\varepsilon_{\rm f}({\bf x},t)]
 \Big/ \langle \varepsilon (t)\rangle_{T>T_0}
   \end{eqnarray}
where $T_0=$ 100 MeV is the temperature constraint. 
This temperature constraint is introduced because the temperature gradients and hence 
the thermal vorticity are very high at the the spectator-participant border, 
where the temperature itself is not that high.  
At the same time, the $\Lambda$ hyperons are abundantly produced 
from the hottest regions of the system. 
Therefore, we applied this temperature constraint on this averaging,  
keeping in mind application the $\Lambda$ polarization. 
The average kinematic vorticity is not that strongly affected by the low-temperature contributions 
of the border regions \cite{Ivanov:2017dff}. This is why this constraint is omitted 
in the case of the kinematic vorticity.  

The above average values of the vorticity are dominated by that in the fragmentation 
regions. Therefore, we use them as an estimates of the vorticity in the fragmentation 
regions, keeping in mind that the true vorticity in the fragmentation 
regions is even higher. 
To estimate the vorticity in the midrapidity region we perform averaging 
over only the center region of the system 
$|x| < R-b/2$, $|y| < R-b/2$ and $|x| < R/\gamma_{cm}$, 
where $R$ is the radius of the Au nucleus and 
$\gamma_{cm}$ is the Lorentz factor associated with the initial nuclear motion in the c.m. frame. 
In fact, this center region is a central layer because  it covers 
the whole participant region in the transverse direction.

Time evolution of the average 
relativistic kinematic and thermal $zx$ vorticities 
calculated in the above described way is displayed in Figs. \ref{fig4} and 
\ref{fig5} for the two considered EoS's. 
The kinematic and thermal vorticities manifest very similar behavior. 
The vorticity of the f-fluid 
averaged with the $\varepsilon_{\rm f}$ weight over the whole system 
is separately presented. 
The f-fluid vorticity is more than an order of magnitude lower than that of 
the baryon-rich fluid, as it is expected from the negligible fraction of the 
total angular momentum accumulated in the f-fluid, see Fig. \ref{fig0}.   
The contribution of the f-fluid only slightly reduces the 
vorticity of the composed matter as compared with the baryon-rich vorticity 
at $t>$ 4 fm/c, when the energy density of the f-fluid becomes small compared 
to the that of the baryon-rich fluid, see Fig. \ref{fig0}. 

The average vorticity in the central region rapidly drops with time. At the early stage 
the average vorticity in the central region practically coincides with the total one 
since this central region includes practically the whole system because of the 
Lorentz contraction of colliding nuclei. Already at $t>$ 2 fm/c,  
the central vorticity is more than 
an order of magnitude lower than the total one. It means the vorticity moves to 
the fragmentation regions.

The vorticity is redistributed in the process of the collision: the vorticity of 
the f-fluid drops, 
the vorticity moves from the central region to the fragmentation ones.  
Nevertheless, the average vorticity of the matter remains approximately constant in the process 
of the collision. The average thermal vorticity even slightly rises with time.

It is worthwhile to mention that the average vorticity displayed in Figs. \ref{fig4} and \ref{fig5}
does not coincide with that of the frozen-out system. 
The freeze-out in the 3FD model is a continuous in time process \cite{3FD,Russkikh:2006aa}, 
as it is illustrated in Figs. \ref{fig1} and \ref{fig2}. 
The actual vorticity of the frozen-out matter can be judged   
from the values of the vorticity field on the freeze-out contour in Figs. \ref{fig1} and \ref{fig2}.
As seen from Figs. \ref{fig1} and \ref{fig2}, average values presented in Figs. \ref{fig4}  and \ref{fig5} 
can be considered as order-of-magnitude estimates of the vorticity averaged over the frozen-out system 
at a fixed time instant. 

\section{Consequences for polarization}
\label{polarization}

The above features of the vorticity imply certain consequences for the observable hyperon polarization. 
The vorticity cannot be presented as a function of the longitudinal rapidity because it is not an 
observable quantity. However, in the self-similar one-dimensional expansion 
of the system the longitudinal space-time rapidity (\ref{eta_s})
equals the kinematic longitudinal rapidity defined in terms of the longitudinal velocity.  
Thus, Fig. \ref{fig1} gives an impression of the rapidity distribution of the vorticity and 
thereby the polarization. 
First of all, this distribution suggests that in semi-central collisions the polarization 
in the fragmentation regions should be at 
least an order of magnitude higher than that observed by the STAR collaboration \cite{STAR:2017ckg} 
in the midrapidity.

The global polarization results from asymmetry of the vortex rings 
which also asymmetrically expand in the transverse direction. The asymmetrical expansion 
results in directed flow of emitted particles. Hence, the global 
polarization in the fragmentation regions should be asymmetrical in the reaction plain with the 
$\Lambda$ and $\bar{\Lambda}$ polarizations correlating with the corresponding directed flow. 

In view of the strong global and even stronger local polarization expected in the 
fragmentation regions one comment is in order. The experimental observation of the global 
polarization \cite{STAR:2017ckg} indicates that a strong spin-orbital coupling of some kind 
is present in the system. This coupling transfers the initially collective angular momentum into 
that accumulated in aligned spins of constituents of the system. If this part 
accumulated in the aligned spins is large, it should affect the collective dynamics of the system 
because the collective angular momentum is respectively reduced. This means the need to 
incorporate the medium polarization into the collective dynamics, e.g. the hydrodynamics 
\cite{Montenegro:2017rbu,Florkowski:2017ruc}. The present 3FD simulations were performed without such feedback.

\section{Summary}
\label{Summary}

Within the 3FD model we have studied vorticity evolution 
in Au+Au collisions at $\sqrt{s_{NN}}=$ 39 GeV. We considered two definitions of the 
vorticity---relativistic kinematic and thermal vorticities---that are relevant in different 
approaches to the hyperon polarization. As found, the kinematic and thermal vorticities
manifest very similar behavior. Moreover, this behavior is very similar within the 
first-order-transition and crossover scenarios.

The vortical fields, both the kinematic and thermal ones, 
are predominately formed at the periphery of the system, 
i.e. at the border between the participant and spectator matter.  
This means that the vorticity is initially located at peripheral rapidities rather than 
at midrapidity. 
Later on, the vortical fields partially spread to the participant and spectator bulk though 
remain concentrated near the border.

A peculiar structure of two vortex rings is formed: one in the target fragmentation region and another in 
the projectile one. The matter rotation is opposite in this two rings. They are 
formed because the matter in the vicinity of the beam axis is stronger decelerated 
than that at the periphery because of thicker matter in the center. 
Thus, the peripheral matter acquires a rotational motion, which  
is  inhomogeneous along the rings  in noncentral collisions. 

These rings are also formed in central collisions. 
In this case the distribution of the vorticity  
is more homogeneous than that in semi-central collisions, but still the vorticity reaches high values. 
These vortex rings are already formed at the early stage of the collision ($t=$ 1 fm/c).  
The  schematic picture of the completely symmetric vortex ring in presented in 
Fig. \ref{fig3a}, corresponding to the exactly central collision.

The average vorticity is responsible for the global polarization of 
the observed $\Lambda$ and $\bar{\Lambda}$ polarizations. 
In the semi-central collisions
the average vorticity in the central region  at $t>$ 2 fm/c,  
when the central region can be associated with the midrapidity region, 
the central vorticity is more than 
an order of magnitude lower than the total one. 
The total vorticity is dominated contributions of 
the fragmentation regions and is produced because of the asymmetry of the above mentioned vortex rings.

The above features of the vorticity imply certain consequences for the observable hyperon polarization. 
First of all, they suggest that in semi-central collisions the global polarization 
in the fragmentation regions should be at 
least an order of magnitude  higher than that observed by the STAR collaboration \cite{STAR:2017ckg} 
in the midrapidity. This polarization should be asymmetrical in the reaction plain: 
the $\Lambda$ and $\bar{\Lambda}$ polarizations should correlate with the corresponding directed flow. 
%


\begin{acknowledgments} 
Fruitful discussions with D.N. Voskresensky are gratefully acknowledged.
This work was carried out using computing resources of the federal collective usage center ``Complex for simulation and data processing for mega-science facilities'' at NRC "Kurchatov Institute", http://ckp.nrcki.ru/.
Y.B.I. was supported by the Russian Science
Foundation, Grant No. 17-12-01427.
A.A.S. was partially supported by  the Ministry of Education and Science of the Russian Federation within  
the Academic Excellence Project of 
the NRNU MEPhI under contract 
No. 02.A03.21.0005. 
\end{acknowledgments}

\end{document}